\documentclass{amsart}
\usepackage{amsfonts,amssymb,amscd,amsmath,enumerate,verbatim,newlfont,calc}
\usepackage{epsfig}
\usepackage{amssymb}
\usepackage{amsfonts}
\usepackage{amsmath,amsthm}
\usepackage{graphicx}
\usepackage{graphics}
\usepackage{rotate}

    \usepackage{amsmath}
    \usepackage{amsthm}
    \usepackage{amssymb}
    \usepackage{latexsym}
\usepackage{amsfonts}

\begin{document}

\title[The effect of an information system on the learning ...]{The effect of an information system on the learning of the space structure}

\author[M.R. MOLAEI]{ {\bf {M.R. MOLAEI}}\\ Mahani Mathematical Research Center\\ Shahid Bahonar University of Kerman, Kerman, Iran
\\ e-mail: mrmolaei@uk.ac.ir}

 \maketitle

\begin{abstract}
In this essay, a general case of information systems contains quantum information systems is considered. By presenting an algorithmic method a new kind of information topology is defined and considered. Continuous maps between two information topological spaces are studied. Moreover, open and compact  information systems  are taken into consideration. It is also proved  that a finite product of compact information systems is a compact information system. Following that, two methods for constructing new open covers for a  class of compact information systems are presented, and information topological entropy for continuous self maps of an information topological space is considered. We show  that information topological entropy is an invariant object under a conjugate relation. Finally, as an applied example, a mathematical model for knowledge spread is introduced.
\end{abstract}
\vspace{1cm}
\maketitle
\section{Introduction}
General theory of systems as a theory of everything in philosophy has a long history in philosophy and history. In this direction one can refer to the works of Leibniz. The works of Bertalanffy and his famous book "General system theory" \cite{BER} is the main reference on the foundations of this topic. In mathematical language a system $Y$ on a non-empty set $X$ is a set of relations on $X$ such that each element of $X$ is in the domain of at least one member of $Y$. In this case the pair $(X,Y)$ is called a systematic space. Information systems are a special class of systematic spaces, in fact in the definition of a systematic space if we replace {\it "relations on $X$"} with {\it "mappings on $X$"}, then it call an information system. More precisely
an information system \cite{G}, \cite{W} is a triple $(X,I, F)$ where
$X$ and $I$ are non-empty sets, and
$F=\{f_{i}~|~f_{i}~is~a~map~on~X,~and~i\in I\}$.  In ordinary information theory, the cardinality of $X$ is smaller than that of real numbers. In the case of quantum information systems, the cardinality of $X$ can be equal to that of real numbers \cite{Dio}.  We ask the reader to pay attention to this point that: in information theory $I$ is a finite set, but in this paper $I$ can be an infinite set. This situation appears in dynamical systems and physical systems in quantum mechanics. In a dynamical system, $I$ is the set of real numbers or any other topological group and in a physical system a member of $F$ is a state or an observable \cite{NAB} which according to the axioms of physical systems, $F$ can be considered as the set of Borel probability measures on $R$. So any quantum information system is also an example of  an information system. \\
In this paper we present an algorithmic method via two binary operations for deducing  essential notions such as openness, compactness, continuity and entropy. These notions have deep backgrounds in physics and ordinary information theory.\\
For a given $(X,I)$ we take $\hat{F}$ as a non-empty subset of $$\{F~|~ (X,I,F)~is~an~information~system\}.$$ We also assume that  $\vee$ and $\wedge $ are two binary operations on $\hat{F}$, and  $(\hat{F}, \vee, \wedge)$ satisfies the following conditions:\\
$(i) ~F\vee (G\vee H)=(F\vee G)\vee H, ~F\wedge (G\wedge H)=(F\wedge G) \wedge H$ for all $F,G, H\in \hat{F}$;\\
$(ii) ~F\vee (F\wedge G)=F\wedge (G\vee F)=F $ for all $F,G\in \hat{F}$;\\
$(iii)~ F\vee F=F\wedge F=F$ for all $F\in \hat{F}$;\\
$(iv) ~ F\wedge (G\vee H)=(F\wedge G)\vee (F\wedge H)$, and $ (G\vee H)\wedge F= (G\wedge F) \vee (H\wedge F)$ for all $F,G, H\in \hat{F}$;\\
$(v) ~  F\vee (G\wedge H)=(F\vee G)\wedge (F\vee H)$, and $(G\wedge H)\vee F= (G\vee F)\wedge (H\vee F)$ for all $F,G, H\in \hat{F}$.\\
 The reader must pay attention to this point that  $\vee$ and $\wedge$ are not necessary commutative, so   $(\hat{F}, \vee, \wedge)$ may not be a distributive lattice \cite{B}.\\
Let  $\hat{G}$ be a subset of $ \hat{F}$. Then $\bigvee \hat{G}$  is the set of $S\in \hat{F}$ with the following properties:\\
$(a)~G=G\wedge S$ for all $G\in \hat{G}$;\\
$(b)$ if there exists $H\in \hat{F}$ such that $G=G\wedge H$ for all $G\in \hat{G}$, then $S=S\wedge H$.\\ An element $S$ with the above properties may not be unique (in the case of existence) (see example 2.3).\\
In the next section a concept of topology via an information system is introduced. Information open and closed sets are studied. Compact information systems and finite product of information systems are considered.\\
In section 4 the notion of information topological entropy for continuous maps on a compact information topological space as an extension of topological entropy \cite{A}, \cite{BO}, \cite{WA}  is introduced. We show that the combination of a map with itself increases the information topological entropy. We introduce a kind of conjugate relation and show that information topological entropy is an invariant object under these kinds of relations.\\ Knowledge spread as an interesting subject in human activities is considered from a mathematical viewpoint in section 5, and  a mathematical model is presented for it. This model is a realistic  example of topological information systems.
\section{A new concept of topology created by an information system}
The name "information topology" has been used previously by Harremoes for a kind of metric spaces created by information systems \cite{HAR}. Here we use of this name for another structure.\\
We assume that $X$, $I$, and $\hat{F}$ have the meanings  determined in the previous section. We choose an information system $(X,I, F)$ and  would like to consider the effect of it for creating the new topological viewpoints.  \\
{\bf Definition 2.1.} An  $F$-topology for $X$ is a subset $\tau_F$ of $\hat{F}$
with the following properties: \\
$(i)~ G=G\wedge F$ for all $G\in \tau_F$;\\
$(ii) $ $F\in \tau_F$, and there is  $O\in \tau_F$ such that $O=O\wedge G$, for all $G\in \tau_F$;\\
$(iii) ~ G\wedge H\in\tau_F$ whenever
$G, H\in \tau_F$;\\
$(iv)$ if $ \emptyset \neq \hat{G}\subseteq \tau_F$, then $~ \bigvee \hat{G}\neq \emptyset $ and $\bigvee \hat{G}\subseteq \tau_F$.\\
$(X, F, \hat{F}, \tau_{F}) $ or briefly $(X, \tau_F ) $ is called an $F$ topological space or an information topological space, and the elements of $\tau_F $ are called  open information systems or $F$ open  systems. \\
 The interior set of  $H\in \hat{F}$ is the set $\bigvee \hat{G}$ where
$\hat{G}= \{ G \in \tau_F ~ | ~ G=G\wedge H\}.$ We denote the interior set of $H$ by $Int H$.\\
{\bf Theorem 2.1.} $H$ is $F$ open if and only if $H\in Int H$.\\
{\bf Proof. } Suppose $\hat{G}= \{ G \in \tau_F ~ | ~ G=G\wedge H\}.$ If $H$ is $F$ open, then $H\in \hat{G}$. Hence, if $H_{0}\in \hat{F}$ and $G=G\wedge H_{0}$ for all $G\in \hat{G}$, then $H=H\wedge H_{0}$. Thus, $H\in \bigvee \hat{G}=Int H$. The converse is obvious, because  $ Int H=\bigvee \hat{G}\subseteq \tau_F$. $\Box $\\
Let $O$ be an object which satisfies axiom $(ii)$ of definition 2.1., and let for given $G\in \hat{F} $, with $G=G\wedge F$, there is a unique $G_{F} \in \hat{F}$ such that $G_{F}=G_{F}\wedge F$, $G\vee G_{F}=G_{F}\vee G=F$, and $G\wedge  G_{F}=G_{F} \wedge G=O$. $G_{F}$ is a mapping of $O$, i.e. if $O$ changes then $G_{F}$ may change. An element $H\in \hat{F}$ is called a closed information system or an $(O,F)$ closed system if $H=H\wedge F$,  and $H_{F}\in \tau_F$.\\
{\bf Theorem 2.2. } If $A$ and $B$ are $(O,F)$ closed, $(A\wedge B)\wedge A_{F}=O$, and $B\vee A_{F}\vee B_{F}=F$, then $A\wedge B$ is $(O,F)$ closed.\\
{\bf Proof. }  $(A\wedge B)\wedge F= A\wedge (B\wedge F)=A\wedge F=F$.\\
$(A\wedge B)_{F}=(A_{F}\vee B_{F})$, for\\
$$ (A\wedge B)\vee (A_{F}\vee B_{F})=( (A\vee A_{F})\vee B_{F})\wedge (B\vee(A_{F}\vee B_{F}))$$ $$= (F\vee B_{F})\wedge (B\vee A_{F}\vee B_{F})= (B\vee B_{F}\vee B_{F})\wedge (B\vee A_{F}\vee B_{F})$$ $$= (B\vee B_{F})\wedge (B\vee A_{F}\vee B_{F})=F\wedge F=F.$$
$$(A\wedge B)\wedge (A_{F}\vee B_{F})=((A\wedge B)\wedge A_{F})\vee ((A\wedge B)\wedge B_{F})$$ $$= O\vee (A\wedge O)=O\vee (A\wedge (A_{F}\wedge A))=O\vee(O\wedge A)=O\vee O=O.$$
Thus $(A\wedge B)_{F}=(A_{F}\vee B_{F})\in \bigvee \{A_{F}, B_{F}\}\subseteq \tau_F.$
 $\Box$\\
 In the next example we show that topological spaces are special cases of information topological spaces.\\
 {\bf Example 2.1.} Suppose $(X,\tau)$ is a topological space. Let $I$ be a singleton, and let $\hat{F}$  be the set $$\{\{f\}~|~ f:X\rightarrow [0,1]~is~a~function~so~that~f(X)~has ~at~most~two~elements.\}.$$ Moreover, let $F=\{\chi _{X}\}$ and  $\tau_{F}=\{\chi_{U}~|~U\in \tau \}$ where $\chi_{.} $ is the characteristic function. If for given $\{f\},\{g\} \in \hat{F}$ we define $\{f\}\wedge \{g\}=\{min\{f,g\}\}$ and $\{f\}\vee \{g\}=\{max\{f,g\}\}$, then $(X,\tau_{F})$ is an $F$ topological space.\\
 The characteristic function of a set gives all the membership information about it, but in many realistic facts, information is limited by some natural limitations such as scientific limitations, social restrictions, and so on. Furthermore, sometimes we need to consider the member of $X$ from different viewpoints. If we assume $I$ is a set which denotes the number of information, and $\mu :X\rightarrow \displaystyle \prod_{i\in I}[0,1]$ denotes  some natural limitations, then $\mu $ is called a $|I|$- dimensional observer of $X$ or a multi-dimensional observer of $X$ where $|I|$ denotes the cardinality of $I$ (which can be infinite), and  $ \displaystyle \prod_{i\in I}[0,1]$ is the Cartesian product of $[0,1]$ \cite{M}. We can denote $\mu $ by $\displaystyle \prod_{i\in I} \mu_{i}$ where $\mu _{i}:X\rightarrow [0,1]$ is a one-dimensional observer of $X$. Thus in this case, if $F=\{\mu_{i}\}$, then $(X, I, F)$ is an information system. We denote the set of $|I|$-dimensional observers of $X$ by $\hat{O}$.  If $\lambda =\displaystyle \prod_{i\in I}\lambda_{i} $ and $\eta =\displaystyle \prod_{i\in I} \eta _{i} $ are two members of $\hat{O}$, then we define $\lambda \vee \eta $ and $\lambda \wedge \eta $ by $(\lambda \vee \eta)_{i}=max \{\lambda_{i} , \eta_{i}\} $ and $(\lambda \wedge \eta)_{i}=min \{\lambda_{i} , \eta_{i}\} $, respectively.\\
 {\bf Example 2.2.} If $\tau _{F}=\{r \mu ~| r\in [0,1] \}$, then $\tau_F$  is an $F$-topology for $X$.\\
 {\bf Example 2.3.} Let  $X= \{ x_1, ... ,x_6 \} $ be the set of six top students in  a university in the six successive years. We denote the marks of them in the lessons  T(topology), A(analysis) and
D(dynamical systems) in table 1. Let $I=\{T,A,D\}$, and let $f_{i}:X\rightarrow [0,1]$ be defined by  $$f_{i}(x_{k})=\frac{1}{100} ~(the ~mark ~of ~x_{k}~ in ~the ~lesson ~i).$$  Then $(X,I,F=\{f_{i}\})$ is an information system. Let $\hat{F}$ be the set $\{\{g_{i}\})~|~g_{i}:X\rightarrow [0,1]~ is~ a~ mapping \}$. We define $\vee $ and $\wedge$ on  $\hat{F}$ by $(g_{1},g_{2},g_{3})\vee (h_{1},h_{2},h_{3})=(max \{g_{1},h_{1}\}, h_{2},h_{3})$ and  $(g_{1},g_{2},g_{3})\wedge (h_{1},h_{2},h_{3})=(min \{g_{1},h_{1}\},g_{2},g_{3})$. If $\tau _{F}=\{(rf_{1}, r,q) ~|~r\in [\frac{1}{2},1], ~and ~~ r,q:X\rightarrow [0,1]~are~two~mappings.\}$, then $(X,\tau _{F})$ is an $F$-topological space.\\ The object $O$ of axiom $(ii)$ of definition 2.1 can be any member of the set \\ $\{(\frac{1}{2}f_{1},r,q)~|~ r,q:X\rightarrow [0,1]~are~two~mappings.\}$.\\
One must pay attention to this point that in example 2.3 $(\hat{F}, \vee, \wedge)$ is not a lattice, because $\vee$, and $ \wedge$ are not commutative. Moreover, if $$\hat{G}=\{(g_{1},g_{2},g_{3}), (h_{1},h_{2},h_{3})\},$$ then $\bigvee \hat{G}=\{(max\{g_{1},h_{1}\}, r, q)~|~ r,q:X\rightarrow [0,1]~are~two~mappings.\},$ so $\bigvee \hat{G}$ is not a singleton. \\
\begin{table}
\begin{center}
\begin{tabular}{|c|c|c|c|}
\hline X& T& A& D \\ \hline
$x_1$& $95$& $80$& $85$ \\
$x_2$& $96$& $93$& $75$ \\
$x_3$& $99$& $87$& $90$ \\
$x_4$& $88$& $87$& $50$ \\
$x_5$& $88$& $85$& $65$ \\
$x_6$& $91$& $88$& $91$ \\
 \hline
\end{tabular}
\end{center}
\caption{An information system}
\end{table}
If $Y$ is a non-empty set and  $f:Y\rightarrow X$ is a mapping, then $(Y,I, Fof=\{f_{i}of\})$ is an information system. Let  $\hat{H}=\{Fof~|~F\in \hat{F}\}$. We define $\vee$ and $\wedge$ on  $\hat{H}$ by $Fof\vee Gof=(F\vee G)of$ and $Fof\wedge Gof=(F\wedge G)of$, and $\tau _{Y}=\{Hof~|~H\in \tau_{F}\}$. Then $(Y,\tau _{Y})$ is an $Fof$ topological space.\\
{\bf Definition 2.2.} Let $(X,I,F, \tau_{F})$ and $(Y,I,G, \tau_{G})$ be two information topological spaces, and let $f:Y\rightarrow X$ be a mapping. Then $f$ is called $(G,F)$ continuous if\\
$(i)$ $Fof=G$ and $Hof\in \tau_{G}$ for all $H\in \tau_{F}$;\\
$(ii)$ if $H,K\in \hat{F}$, then $Hof, Kof \in \hat{G}$, $Hof\vee Kof=(H\vee K)of$ and $Hof\wedge Kof=(H\wedge K)of$.\\
The straightforward calculations imply to the next theorem.\\
{\bf Theorem 2.3.} If $(X,I,F, \tau_{F})$, $(Y,I,G, \tau_{G})$ and $(Z,I,H,\tau_{H})$ are three information topological spaces,  $f:Y\rightarrow X$, $g:Z\rightarrow Y$ are $(G,F)$ continuous, and $(H,G)$ continuous, then $fog: Z\rightarrow X$ is  $(H,F)$ continuous.\\
\section{Compact information systems}
Suppose $(X,I,F, \tau_{F})$ is an $F$ topological space where $\tau_{F}\subseteq  \hat{F}$. Let  $H\in \hat{F}$, and $H=H\wedge F$. An open cover for $H$ is a subset  $\hat{H}$ of $\tau_{F}$ such that $H=H_{1}\wedge F$, and $H_{1}\in \bigvee \hat{H}$.\\
{\bf Definition 3.1.} An element $H$ of $\hat{F}$ with the property $H=H\wedge F$ is called a compact information system or an $F$ compact system if it has an open cover and  each open cover of $H$ has a finite subset which is also an open cover for $H$.\\
{\bf Theorem 3.1.} If $H$ and $K$ are two  $F$ compact systems, and if any open cover of $H\vee K$ is an open cover of $H$ and $K$, then $H\vee K$ is an $F$ compact system.\\
{\bf Lemma 3.1.} If $\hat{H}\subseteq \hat{F}$, and  $\hat{K}\subseteq \hat{F}$, then  $\{H\vee K~|~ H\in \bigvee \hat{H}, ~and ~K\in \bigvee \hat{K}\}\subseteq \bigvee (\hat{H}\cup \hat{K})$.\\
{\bf Proof. } Let $H\in \bigvee \hat{H}, ~and ~K\in \bigvee \hat{K}$. Then for $G\in \hat{H}$ we have
$$G\wedge (H\vee K)= (G\wedge H)\vee (G\wedge K)= G\vee (G\wedge K)= G.$$
Moreover, for $G\in \hat{K}$ we have
$$ G\wedge (H\vee K)= (G\wedge H)\vee (G\wedge K)=(G\wedge H)\vee G$$ $$=(G\vee G)\wedge (H\vee G)=G\wedge (H\vee G)=G.$$
Thus, $ G\wedge (H\vee K)=G$ for all $G\in (\hat{H}\cup \hat{K})$. If there is $D$ such that  $G=G\wedge D$ for all $G\in  (\hat{H}\cup \hat{K})$, then $$(H\vee K)\wedge D=(H\wedge D)\vee (K\wedge D)= H\vee K.$$
Hence, $ H\vee K \in \bigvee (\hat{H}\cup \hat{K})$. $\Box$\\
{\bf Proof of theorem 3.1.} Let $\hat{H}$ and $\hat{K}$ be open covers of $H$ and $K$ respectively. Furthermore let $H_{1}\in \bigvee \hat{H}$, $H_{2}\in \bigvee \hat{K}$, $H=H_{1}\wedge F$ and $K=K_{1}\wedge F$. Then lemma 3.1 implies $H_{1}\vee K_{1}\in \bigvee (\hat{H}\cup \hat{K})$. Moreover, we have $$(H_{1}\vee K_{1})\wedge F=(H_{1}\wedge F)\vee (K_{1}\wedge F)=H\vee K.$$ Thus $\hat{H} \cup \hat{K}$ is an open cover for $H\vee K$. Now let $\hat{G}$ be an open cover for $H\vee K$. Then there exist finite subcovers $\hat{G_{H}}$ and $\hat{G_{K}}$ of $\hat{G}$ with $H_{1}\in \bigvee \hat{G_{H}}$, $K_{1}\in  \bigvee \hat{G_{K}}$, $H=H_{1}\wedge F$ and $K=K_{1}\wedge F.$
$\hat{G_{H}}\cup \hat{G_{K}}$ is a finite subcover of $\hat{G}$, and $H\vee K=(H_{1}\vee K_{1})\wedge F$. Thus  $H\vee K$ is $F$ compact. $\Box$\\
Now we assume that $(Y, G, \hat{G}, \tau_{G})$ is a topological information system,  $f:Y\rightarrow X$ is a $(G,F)$ continuous map. Moreover assume that $\hat{G}=\{Hof ~: ~H\in \hat{F}\}$ and if $H_{1}of=H_{2}of$, then $H_{1}=H_{2}$. With these assumptions we have the next theorem.\\
{\bf Theorem 3.2.} If $Dof\in \hat{G}$ is $G$ compact, then $D$ is $F$ compact.\\
{\bf Proof. } Let $\hat{Q}=\{Hof~|~H\in \hat{D}\}$ be an open cover for $Dof$. Then there is $D_{1}of \in \bigvee \hat{Q}$ such that $Dof= D_{1}of\wedge Fof=(D_{1}\wedge F)of$. So $D=D_{1}\wedge F$. We show that $D_{1}\in \bigvee \hat{D}$. Since $Hof=Hof \wedge D_{1}of=(H\wedge D_{1})of$, then $H=H\wedge D_{1}$ for all $H\in \hat{D}$. If $P\in \hat{F} $ and $H=H\wedge P$ for all $H\in \hat{D}$, then $Hof=(H\wedge P)of$ for all $Hof\in \hat{Q}$. Thus $D_{1}of=D_{1}of\wedge Pof=(D_{1}\wedge P)of$. Hence $D_{1}=D_{1}\wedge P$. Consequently $\hat{D}$ is an open cover for $D$.\\
If $\hat{D}$ is an open cover for $D$, then it creates an open cover $\hat{Q}$ for $Dof$. So it has a finite subcover. This finite subcover corresponds to a finite subcover of $\hat{D}$ for $D$. Thus $D$ is $F$ compact. $\Box$\\
Let $A$ be an index set, and let for $a\in A$, $(X_{a}, F_{a}=\{f_{i}^{a}\}, I, \hat{F}_{a}, \tau _{F_{a}})$ be an information topological space. We define $\displaystyle \prod _{a\in A}F_{a}$ and $\displaystyle \prod _{a\in A}\hat{F}_{a}$ by $\{\{\displaystyle \prod _{a\in A}f_{i}^{a}\} ~|~i\in I\}$ and $\{\displaystyle \prod _{a\in A}G_{a}~|~ G_{a}\in \hat{F}_{a}\}$ respectively. We define $\wedge $ and $\vee $ on $\displaystyle \prod _{a\in A}\hat{F}_{a}$ by $$(\prod _{a\in A}G_{a})\wedge (\prod _{a\in A}H_{a})= \prod _{a\in A}(G_{a}\wedge H_{a}), ~and ~ (\prod _{a\in A}G_{a})\vee (\prod _{a\in A}H_{a})= \prod _{a\in A}(G_{a}\vee H_{a}).$$
If we define $\displaystyle \prod _{a\in A}\tau _{F_{a}}$ by $\{\displaystyle \prod _{a\in A}H_{a} ~ : ~ H_{a}\in \tau _{F_{a}}\}$, then \\ $(\displaystyle \prod _{a\in A}X_{a},\displaystyle \prod _{a\in A}F_{a}, I,\displaystyle \prod _{a\in A}\hat{F}_{a},\displaystyle \prod _{a\in A}\tau _{F_{a}})$ is an information topological space, which we call a product information topological space.\\
{\bf Theorem 3.3. } If $A$ is a finite index set, and $(X_{a}, \tau _{F_{a}})$ is an $F_{a}$ compact system, then $(\displaystyle \prod _{a\in A}X_{a},\displaystyle \prod _{a\in A}\tau _{F_{a}})$ is a $\displaystyle \prod _{a\in A}F_{a}$ compact system.\\
{\bf Proof.}
For $a\in A$, $\hat{G}_{a}$ is an open cover for $F_{a}$ if and only if $F_{a}=H_{a}\wedge F_{a}$, and $H_{a}\in \bigvee \hat{G} _{a}$. This is equivalent to
$\displaystyle \prod _{a\in A}F_{a}=(\displaystyle \prod _{a\in A}H_{a})\wedge (\displaystyle \prod _{a\in A}F_{a})$, and $\displaystyle \prod _{a\in A}H_{a}\in \bigvee (\displaystyle \prod _{a\in A}\hat{G}_{a})$.  Consequently, for $a\in A$, $\hat{G}_{a}$ is an open cover for $F_{a}$ if and only if $\displaystyle \prod _{a\in A}\hat{G}_{a} $ is an open cover for $\displaystyle \prod _{a\in A}{F}_{a}$.\\  Since each $F_{a}$ has an open cover, then $\displaystyle \prod _{a\in A}{F}_{a}$ has an open cover. Now let $\displaystyle \prod _{a\in A}\hat{G} _{a}$ be an open cover for $\displaystyle \prod _{a\in A}F_{a}$. Then for $a\in A$, $\hat{G} _{a}$ has a finite subcover  $\hat{H} _{a}$ for $F_{a}$. Hence $\displaystyle \prod _{a\in A}\hat{H} _{a}$ is a finite open subcover for $\displaystyle \prod _{a\in A}F_{a}$. As a result, it is a compact information system. $\Box$\\
\section{Information topological entropy}
 In this section we assume that $(X,I,F, \hat{F}, \tau_{F})$ is an information  topological space, and $(\hat{F}, \wedge)$ is commutative.\\
 {\bf Theorem 4.1.} If $\hat{H}$ and $\hat{G}$ are open covers for $H\in \hat{F}$, then $\hat{A}=\{E\wedge G~|~E,G\in \hat{H}\cup \hat{G}\}$ is an open cover for $H$.\\
 {\bf Proof.} Clearly $\hat{A}\subseteq \tau_{F}$. Since $H=H_{1}\wedge F$, $H_{1}\in \bigvee \hat{H}$, $G=G_{1}\wedge F$ and $G_{1}\in \bigvee \hat{G}$, then $$(H_{1}\vee G_{1})\wedge F=(H_{1}\wedge F)\vee (G_{1}\wedge F)=H\vee H=H.$$  To complete the proof we must show that $\hat{H_{1}}\vee \hat{G_{1}} \in \bigvee \hat{A}.$
 For given $E\wedge G\in \hat{A}$ we have $$(E\wedge G)\wedge (H_{1}\vee G_{1})=(E\wedge G\wedge H_{1})\vee (E\wedge G \wedge G_{1})$$
 $$ =\left\{ \begin{array}{lll} (E\wedge G)\vee (E\wedge G)=E\wedge G & if ~ E\in \hat{H},~ G\in \hat{G}\\
 (E\wedge G)\vee (E\wedge G \wedge G_{1})=E\wedge G & (by ~ axiom ~(ii))~if ~ E,G\in \hat{H} \\ ((E\wedge G)\wedge H_{1})\vee (E\wedge G)=\\ ((E\wedge G)\vee (E\wedge G))\wedge (H_{1}\vee (E\wedge G))=\\
  (E\wedge G)\wedge (H_{1}\vee (E\wedge G))=E\wedge G& (by ~ axiom ~(ii))~if~ E,G\in \hat{G} \end{array}.\right. $$
  If for some $K\in \hat{F}$, $$E\wedge G\wedge K=E\wedge G~ for ~all ~ E\wedge G\in \hat{A},$$ then $$E\wedge K=E\wedge E \wedge K=E\wedge E=E~ for ~all~ E\in \hat{H},$$
  and $$G\wedge K=G\wedge G \wedge K=G\wedge G=G~ for ~all~ G\in \hat{G}.$$ Thus $H_{1}=H_{1}\wedge K$ and $G_{1}=G_{1}\wedge K$. Hence $(H_{1}\vee G_{1})\wedge K=(H_{1}\wedge K)\vee (G_{1}\wedge K)=H_{1}\vee G_{1}.$ Consequently ${H_{1}}\vee {G_{1}} \in \bigvee \hat{A}.$ $\Box$\\
  {\bf Theorem 4.2.} Let $\hat{H}$, and $\hat{G}$ be open covers for $H\in \hat{F}$. Then $\hat{H}\sqcup \hat{G}=\{E\wedge G ~|~ E\in \hat{H},~and~ G\in \hat{G}\}$ is an open cover for $H$.\\
  {\bf Proof.} We have $H=H_{1}\wedge F$, $H=G_{1}\wedge F$, $H_{1}\in \bigvee \hat{H}$, and $G_{1}\in \bigvee \hat{G}$. Thus $$(H_{1}\wedge G_{1})\wedge F=H_{1}\wedge (G_{1}\wedge F)=H_{1}\wedge H=H_{1}\wedge (H_{1}\wedge F)$$ $$=(H_{1}\wedge H_{1})\wedge F=H_{1}\wedge F=H.$$
  We show that $H_{1}\wedge G_{1}\in \bigvee (\hat{H}\sqcup \hat{G})$. Let $E\wedge G \in \hat{H}\sqcup \hat{G}$ be given. Then $$E\wedge G\wedge H_{1}\wedge G_{1}=E\wedge H_{1}\wedge G \wedge G_{1}=E\wedge G.$$ Suppose there is $D\in \hat{F}$ such that $E\wedge G\wedge D=D$ for all $E\wedge G\in \hat{H}\sqcup \hat{G}$. We choose an $E\wedge G\in \hat{H}\sqcup \hat{G}$, and we have $$H_{1}\wedge G_{1}\wedge D=H_{1}\wedge G_{1}\wedge (E\wedge G)\wedge D=(E\wedge H_{1})\wedge (G\wedge G_{1}) \wedge D=E\wedge G\wedge D=D.$$ Thus $H_{1}\wedge G_{1}\in \bigvee (\hat{H}\sqcup \hat{G})$. $\Box $\\
  If $F$ is a compact system and $\hat{H}$ is an open cover of it, then we denote the cardinality of a subcover of $\hat{H}$ with the smallest cardinality by $N(\hat{H})$.\\
  We denote the entropy of $\hat{H}$ by $E(\hat{H}) $ and  define it by $E(\hat{H})=log(N(\hat{H})$.\\
  {\bf Theorem 4.3.} Let $\hat{H}$ and $\hat{G}$ be open covers for $F$. Then $E(\hat{H}\sqcup \hat{G})\leq E(\hat{H})+E(\hat{G})$.\\
  {\bf Proof.} Let $A=\{H_{1},..., H_{m}\}\subseteq \hat{H}$ and $B=\{G_{1},...,G_{n}\}\subseteq \hat{G}$ be covers of $F$ with the smallest cardinality. Then $A\sqcup B\subseteq \hat{H} \sqcup \hat{G}$ is a cover of $F$. Thus $$E(\hat{H}\sqcup \hat{G})\leq log(card(A\sqcup B))=log(mn)=logm + log n= E(\hat{H})+E(\hat{G}). ~ \Box $$
  {\bf Theorem 4.4.} If $\hat{H}$ is an open cover for $F$, then $\hat{H}of$ is also an open cover for $Fof$.\\
  {\bf Proof.} There is $H_{ 1}\in \hat{F}$ such that $F=H_{1}\wedge F$, and $H_{1}\in \bigvee \hat{H}$. The straightforward calculations imply $Fof=H_{1}of\wedge Fof$, and $H_{1}of\in \bigvee \hat{H}of$. $\Box$\\
  {\bf Theorem 4.5.} If $Fof=F$, then $\lim_{n\rightarrow \infty}\frac{1}{n}E(\displaystyle \bigsqcup_{i=0}^{n-1}\hat{H}of^{i})$ exists.\\
  {\bf Proof.} Let $a_{n}=E(\displaystyle \bigsqcup_{i=0}^{n-1}\hat{H}of^{i})$. Then $$a_{n+k}=E(\displaystyle \bigsqcup_{i=0}^{n+k-1}\hat{H}of^{i})=E((\displaystyle \bigsqcup_{i=0}^{n-1}\hat{H}of^{i})\sqcup (\displaystyle \bigsqcup_{i=n}^{n+k-1}\hat{H}of^{i})$$ $$\leq E(\displaystyle \bigsqcup_{i=0}^{n-1}\hat{H}of^{i})+ E((\displaystyle \bigsqcup_{j=0}^{k-1}\hat{H}of^{j})of^{n})\leq a_{n}+a_{k}.$$
  Thus the sequence $\{a_{n}\}$ is a sub-additive sequence. Accordingly,
   $\lim \frac{a_{n}}{n}=inf \frac{a_{n}}{n}.$ $\Box$\\
  By the assumptions of theorem 4.5, we denote $\lim_{n\rightarrow \infty}\frac{1}{n}E(\displaystyle \bigsqcup_{i=0}^{n-1}\hat{H}of^{i})$  by $e(f,\hat{H},F)$, and we call it the entropy of $f$ relative to $(\hat{H},F)$. \\ We denote $\displaystyle \sup_{\hat{H}}e(f,\hat{H},F)$ by $e(f,F)$ and  call it the information topological entropy or $F$ topological  entropy of $f$.\\
  {\bf Theorem 4.6.} If $Fof=F$, and $k\in N$, then $e(f^{k},F)\geq ke(f,F).$\\
  {\bf Proof.} If $\hat{H}$ is an open cover for $F$, then $\displaystyle \bigsqcup_{i=0}^{n-1}\hat{H}of^{ki}\subseteq \displaystyle \bigsqcup_{i=0}^{kn-1}\hat{H}of^{i} $. As a result, $$a_{n}(f^{k})=E(\displaystyle \bigsqcup_{i=0}^{n-1}\hat{H}of^{ki})\geq E(\displaystyle \bigsqcup_{i=0}^{kn-1}\hat{H}of^{i})=a_{kn}(f). $$ Hence $$k\frac{a_{kn}(f)}{kn}=\frac{a_{kn}(f)}{n}\leq \frac{a_{n}(f^{k})}{n}.$$ By tending $n$ to infinity, we have $ke(f,\hat{H},F)\leq e(f^{k},\hat{H},F).$ Thus $e(f^{k},F)\geq ke(f,F).$ $\Box$\\
Now we assume that $(Y,I, G, \hat{G}, \tau_{G})$ is an information topological space and $g:Y\rightarrow Y$ is a $(G,G)$ continuous map. We also assume that $h:Y\rightarrow X$ is a one to one and onto $(G,F)$ continuous map and $h^{-1}:X \rightarrow Y$ is an $(F,G)$ continuous map (in this case we say that $h$ is a $(G,F)$ homeomorphism). Moreover, we assume that $hog=foh$. With these assumptions we have the next theorem.\\
{\bf Theorem 4.7.} $e(g,G)=e(f,F)$.\\
{\bf Proof.} Since $(\hat{F},\wedge ) $ is commutative, and $\hat{F}oh=\hat{G}$, then $(\hat{G}, \wedge )$ is commutative. We have $$Gog=Goh^{-1}ofoh=Fofoh=Foh=G.$$ Let $\hat{H}$ be an open cover for $F$. Then $\hat{H}ofoh$ is an open cover for $\hat{G}$. In fact the proof of theorem 4.3 implies there is $\hat{H_{1}}\in \hat{F}$ such that $F=(H_{1}of)\wedge F$, and $H_{1}of \in \displaystyle \bigvee(\hat{H}of)$. Since $h$ is a $(G,f)$ homeomorphism, then $\hat{H}_{1}ofoh\in \hat{G}$, and $G=Foh=(H_{1}ofoh)\wedge Foh=(H_{1}ofoh)\wedge G.$ We show that $H_{1}ofoh\in \displaystyle \bigvee (\hat{H}ofoh).$\\ For given $Hofoh\in \hat{H}ofoh$ we have $$(Hofoh)\wedge (H_{1}ofoh)=(H\wedge H_{1})ofoh=Hofoh.$$ If there is $S\in \hat{G}$ such that $(Hofoh)\wedge S=Hofoh$ for all $Hofoh\in \hat{H}ofoh$, then $Soh^{-1}\in \hat{F}$, and $(Hof)\wedge (Soh^{-1})=Hof$ for all $Hof\in \hat{H}of.$ Thus $(H_{1}of)\wedge (Soh^{-1})=H_{1}of.$ Hence $(H_{1}ofoh)\wedge (Soh^{-1}oh)=H_{1}ofoh.$ So $(H_{1}ofoh)\wedge S=H_{1}ofoh.$ This prove that $H_{1}ofoh\in \displaystyle \bigvee(\hat{H}ofoh).$\\
We have $$E(\displaystyle \bigsqcup_{i=0}^{n-1}\hat{H}of^{i})=E(\displaystyle \bigsqcup_{i=0}^{n-1}\hat{H}of^{i}oh)=E(\displaystyle \bigsqcup_{i=0}^{n-1}\hat{H}ohog^{i}).$$
Thus $e(f,\hat{H},F)=e(g,\hat{H}oh,G)$. Since there is a one to one correspondence between open covers of $F$ and open covers of $G$ via the mapping $\hat{H}\mapsto \hat{H} oh$, then $e(f,F)=e(g,G)$. $\Box$\\
\section{Modeling of knowledge spread as an information system}
Knowledge spread or extension of knowledge is one of the most important topics in  human societies.
 Knowledge spreads in a society by different methods such as books, teachers,  magazines, internet, and so on. Suppose $X$ is the set of methods for extending knowledge, and  the probability that
a knowledge fact transforms in a small time interval $\Delta t$ by a method $x\in X$ is equal to $\gamma (x) \Delta t +O((\Delta t)^{2})$, where $\gamma :X\rightarrow [0,1]$ is a mapping. We assume $N(t)$ is the number of population of a society at time $t$ who are interested to receive knowledge facts. Let $I$ be a bounded interval in $R$ with $0\in I$, and let $M=sup\{N(t)~|~t\in I\}$. $D^{\gamma}(N(t))(x)$ denotes  the expected number of population at time $t$ who receive a knowledge fact via a method $x\in X$. $p_{n}^{\gamma,x}(t)$ is the probability that exactly $n$ individuals receive a knowledge fact at time $t$ via the method $x$. Thus $D^{\gamma}(N(t))(x)=\displaystyle \Sigma _{n=0}^{M} np_{n}^{\gamma,x}(t).$  Let $F^{\gamma }=\{D^{\gamma}(N(t))~|~t\in I\}$. Then $(X,I,F^{\gamma })$ is an information system.  To determine $F^{\gamma}$, we must determine the evolution of  $p_{n}^{\gamma,x}(t)$.\\
For a fixed method we know that the probability of an individual  receiving a knowledge fact at time $t+\Delta t $ is equal to the probability of an individual receiving a knowledge fact at time $t$ or nobody receiving that fact at time $t$, but one person receiving a knowledge fact at a time in the interval  $(t,t+ \Delta t )$. Hence $p_{1}^{\gamma,x}(t)=p_{1}^{\gamma,x}(t)+\gamma (x) \Delta t$. If $n>1$, then the probability of $n$ individuals  receiving a knowledge fact at time $t+\Delta t $ is equal to the probability of $n$ individual receiving a knowledge fact at time $t$ or $n-1$ individuals receiving that fact at time $t$, and they  extend it with the probability $(n-1)\gamma(x)$ to another person in the interval time $(t,t+ \Delta t )$. Thus $p_{n}^{\gamma,x}(t+\Delta t)=(n-1)p_{n-1}^{\gamma,x}(t)\gamma (x)\Delta t+p_{n}^{\gamma,x}(t)+O((\Delta t)^{2})$.\\
 Hence $$p_{n}^{\gamma,x}(t+\Delta t)= \left\{ \begin{array}{ll} (n-1)p_{n-1}^{\gamma,x}(t)\gamma (x)\Delta t+p_{n}^{\gamma,x}(t)+O((\Delta t)^{2}) & if ~ 1<n\leq M\\ p_{1}^{\gamma,x}(t)+\gamma (x) \Delta t & if ~n=1\\ p_{0}^{\gamma,x}(t)+(1-\gamma (x)) \Delta t & if ~ n=0\end{array}.\right.$$ By tending $\Delta t$ to zero, we find the following system of differential equations
 $$\frac{dp_{n}^{\gamma,x}(t)}{dt}=\left\{ \begin{array}{ll}(n-1)\gamma (x) p_{n-1}^{\gamma,x}(t)& if ~ 1<n\leq M\\ \gamma (x)  & if ~n=1\\(1-\gamma (x))  & if ~ n=0\end{array}.\right.$$
    As a result $$p_{n}^{\gamma,x}(t)=\left\{ \begin{array}{ll}(n-1)!\gamma (x)^{n}\frac{t^{n}}{n!}+(n-1)!\gamma (x)^{n-1}p_{1}^{x}(0)\frac{t^{n-1}}{(n-1)!} + ... +\\(n-1)!\gamma (x) p_{n-1}^{\gamma,x}(0)t+p_{n}^{\gamma,x}(0)& if ~ 1<n\leq M\\ \gamma (x) t+p_{1}^{\gamma,x}(0)  & if ~n=1\\(1-\gamma (x))t+ p_{0}^{\gamma,x}(0) & if ~ n=0\end{array}.\right.$$
   As we see the velocity of knowledge spread depends on $\gamma $. $\gamma $ determines the interest to knowledge in a society. For example if we associate a real number to each method and take $\gamma(x)=\frac{1+3x+[-3x]}{100}$ (see figure 1), then we have illustrated the graphs of $p_{2}^{\gamma,4}(.),~ p_{2}^{\gamma,4.9}(.),$ and $p_{2}^{\gamma,4.5}(.)$ in figure 2. We have sketched $p_{2}^{\gamma,.}(.)$ in figure 3. \\
   \begin{figure}[]
\centering
\includegraphics[width=6cm, height=6cm]{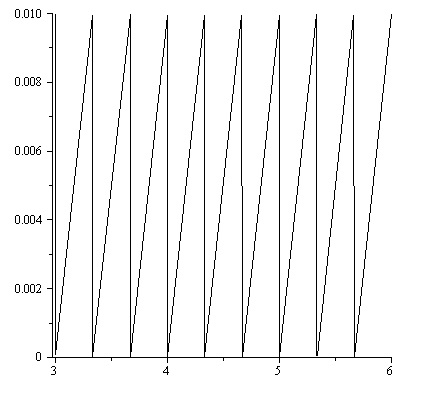}
\caption{The graph of $\gamma .$}
\label{}
\end{figure}
\begin{figure}[]
\centering
\includegraphics[width=6cm, height=6cm]{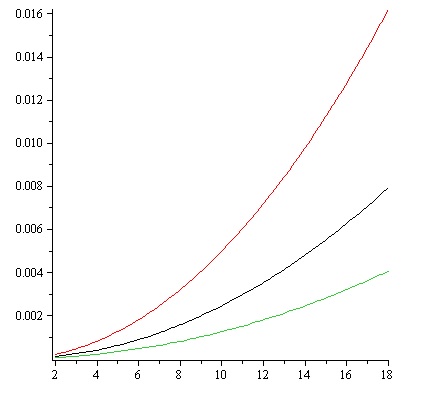}
\caption{Black, green, and red colors denote the graphs of $p_{2}^{\gamma,4.9}(.),~ p_{2}^{\gamma,4.5}(.),~p_{2}^{\gamma,4}(.)$ respectively.}
\label{}
\end{figure}
\begin{figure}[]
\centering
\includegraphics[width=6cm, height=6cm]{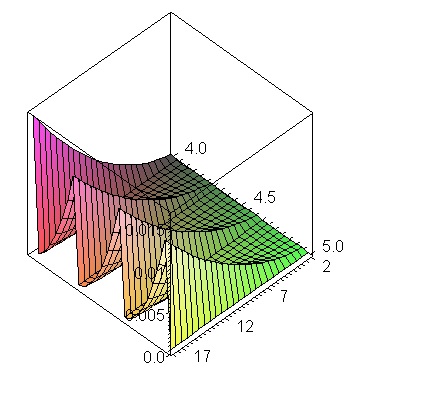}
\caption{The graph of $p_{2}^{\gamma,.}(.)$}
\label{}
\end{figure}

   Suppose $I$ and $X$ are fixed. Take    $$\hat{F}=\{F^{\gamma}~|~ \gamma :X\rightarrow [0,1]~is~a~mapping\}. $$ For given $F^{\gamma}, F^{\delta}\in \hat{F}$ we define  $F^{\gamma}\wedge F^{\delta}$ by $F^{\eta}$, where $\eta:  X \rightarrow [0,1]$ is defined by $\eta(x)=min\{\gamma (x), \delta (x)\}. $ We also define  $F^{\gamma}\vee F^{\delta}$ by $F^{\lambda}$, where $\lambda :X\rightarrow [0,1]$ is the mapping $\lambda (x)=max \{\gamma (x), \delta (x)\}$. Let $\mu :X\rightarrow [0,1]$ be the constant function $\mu (x)=\frac{1}{M}$, and let $F=F^{\mu }$. Then $$\tau _{F}=\{F^{\frac{1}{M}\gamma}~|~ \gamma:X\rightarrow [0,1]~is~a~mapping\}$$ is an $F$-topology for $X$. Although $X$ is finite,  $F$ is not a compact information system.\\
   If we take $$\tau _{F}=\{F^{\frac{1}{M}\gamma_{x}}~|~ x\in X ~and ~ \gamma_{x}:X\rightarrow [0,1]~can~be~any~map~with~\gamma_{x}(x)=1\},$$ then the finiteness of $X$ and the structure of $\tau _{F}$ imply $F$ is a compact information system. Since $X$ is finite, we denote it by $X=\{x_{1},..., x_{k}\}$. Let $f:X\rightarrow X$ be the shift map, i.e. $f(x_{i})=x_{i+1}$ for $i=1,..., k-1$ and $f(x_{k})=x_{1}$. Then $F^{\frac{1}{M}\gamma _{x_{i}}}of=F^{\frac{1}{M}\gamma _{x_{i-1}}}$ for $i=2,..., k$ and $F^{\frac{1}{M}\gamma _{x_{1}}}of=F^{\frac{1}{M}\gamma _{x_{k}}}$. Hence $e(f,F)=0$. \\ With the above $F$ topology, one can apply theorem 4.4 and prove that if $f:X\rightarrow X$ is an arbitrary $(F,F)$ continuous map, then $e(f,F)=0$.\\
   Now we assume that $X$ is an infinite countable set, and we denote it by $X=\{...,x_{-2},x_{-1},x_{0},x_{1},x_{2},...\}$. Let $k$ be a fixed natural number, and let $\theta :X\rightarrow [0,1]$ be a fixed mapping. Besides let $F=F^{\theta }$, and $$\tau_{F}=\{F^{\gamma_{i}}~|~ \gamma_{i}:X\rightarrow [0,1]~is~any~map~with $$ $$~\gamma_{i}(x_{j})=\theta (x_{j})~for~j\notin \{-i+1,-i+2,...,i-2,i-1\}~and~i> k \}.$$
   If $f:X\rightarrow X$ is the shift map $f(x_{i})=x_{i+1}$, then $\gamma_{i}of=\gamma_{i+1}.$ In this case the straightforward calculations imply $e(f,F^{\theta })=log(2k).$
\section{Conclusion}
Information systems as  essential  objects have a main role in our living structures. We determine their effect on our space viewpoints. In fact we determine their effect on the notion of topology,  continuity, and entropy. We show that a class of  systems with two binary operations can determine the space structure. This means that a set of theoretical axioms with two binary operations  which satisfy the properties which are mentioned in the second page of this article can determine a new viewpoint of the space.

\end{document}